\documentclass[12pt]{article}
\usepackage{amsmath}
\usepackage{cite}
\usepackage{color}
\usepackage{amssymb}
\usepackage{amsfonts}
\usepackage{booktabs}
\usepackage{multirow}
\usepackage{pdflscape}
\usepackage{gensymb}
\usepackage{graphicx}
\usepackage{bigstrut}
\usepackage{pifont}
\usepackage{enumerate}
\usepackage[small,bf]{caption}

\newcommand\pubnumber{NuPhys2017-Penedo}
\newcommand\pubdate{\today}

\def\Title#1{\begin{center} {\Large #1 } \end{center}}
\def\Author#1{\begin{center}{ \sc #1} \end{center}}
\def\Address#1{\begin{center}{ \it #1} \end{center}}

\newcommand\pubblock{\rightline{\begin{tabular}{l} \pubnumber\\
         \pubdate  \end{tabular}}}
\newenvironment{Abstract}{\begin{quotation}  }{\end{quotation}}
\newenvironment{Presented}{\begin{quotation} \begin{center} 
             PRESENTED AT\end{center}\bigskip 
      \begin{center}\begin{large}}{\end{large}\end{center} \end{quotation}}
\def\Acknowledgements{\bigskip  \bigskip \begin{center} \begin{large}
             \bf ACKNOWLEDGEMENTS \end{large}\end{center}}

\begin{document}
\begin{titlepage}
\pubblock

\vfill
\Title{
Neutrino Mixing and Leptonic CP Violation\\[2mm] from $S_4$ and Generalised CP Symmetries}
\vfill
\Author{J. T. Penedo$^{\,a,}$\footnote{Poster presenter},
S. T. Petcov$^{\,a,b,}$\footnote{Also at: Institute of Nuclear Research and
Nuclear Energy, Bulgarian Academy of Sciences, 1784 Sofia, Bulgaria},
A. V. Titov$^{\,c}$}
\Address{$^{a\,}$SISSA/INFN, Via Bonomea 265, 34136 Trieste, Italy\\
$^{b\,}$Kavli IPMU (WPI), University of Tokyo, 5-1-5 Kashiwanoha,\\ 277-8583 Kashiwa, Japan\\
$^{c\,}$Institute for Particle Physics Phenomenology, Department of Physics,\\ Durham University, South Road, Durham DH1 3LE, United Kingdom
}
\vfill
\begin{Abstract}
We consider a class of models of neutrino mixing 
with $S_4$ flavour symmetry and
generalised CP symmetry, broken to 
$Z_2$ and $Z_2 \times {\rm CP}$ residual symmetries in 
the charged lepton and neutrino sectors, respectively.
In this scheme, and up to discrete ambiguities, the neutrino mixing matrix 
is determined by two angles and one phase. 
We classify the phenomenologically viable mixing patterns,
deriving predictions for the Dirac and Majorana CPV phases
and for the effective Majorana mass in neutrinoless double beta decay.
\end{Abstract}
\vfill
\begin{Presented}
NuPhys2017, Prospects in Neutrino Physics\\[1mm]
Barbican Centre, London, UK,  December 20--22, 2017
\end{Presented}
\vfill
\end{titlepage}
\def\thefootnote{\fnsymbol{footnote}}
\setcounter{footnote}{0}

\section{Introduction}

The patterns that have emerged from neutrino oscillation data
in recent years (see e.g.~\cite{Olive:2016xmw})
offer a potential window into the origins of flavour.
Extensions of the Standard Model with 
non-Abelian discrete flavour symmetries (see e.g.~\cite{Altarelli:2010gt,Ishimori:2010au})
have been considered extensively in attempts to 
understand the flavour problem.
While the flavour symmetry
may determine the neutrino mixing angles 
and/or the Dirac phase, a
generalised CP (gCP) symmetry \cite{Branco:1986gr},
implemented in a consistent way \cite{Feruglio:2012cw,Holthausen:2012dk},
allows also to constrain also Majorana CP violating (CPV) phases.

In the flavour+gCP approach, a fundamental symmetry described by a group $G_\text{CP} = G_f \rtimes H_\text{CP}$ is assumed to be realised at some high-energy scale and to be broken at lower
energies to residual symmetries $G_e$ and $G_\nu$, in the charged-lepton and neutrino sectors, respectively. Here, $G_f$ is a flavour symmetry group admitting a 3D irreducible representation $\rho$, while $H_\text{CP}$ denotes a group of gCP transformations.

The present contribution is based on the work of Ref.~\cite{Penedo:2017vtf}, in which we take $G_f = S_4$, $G_e = Z_2$ and $G_\nu = Z_2 \times \text{CP}$. After briefly reviewing our approach, we summarise the phenomenological consequences of this simple breaking pattern.

\section{Framework}
The residual flavour symmetries are associated to the group elements $g_e$ and $g_\nu$.
The residual gCP transformation in the neutrino sector is instead
described by a matrix $X_\nu$ in flavour space.
These residual symmetries constrain the charged-lepton and neutrino mass matrices, $M_e$ and $M_\nu$, which satisfy
\begin{align}
\rho(g_e)^\dagger M_e M_e^\dagger\, \rho(g_e) = M_e M_e^\dagger\,,
\quad
\rho(g_\nu)^T M_\nu\, \rho(g_\nu) = M_\nu\,,
\quad
X_\nu^T M_\nu\, X_\nu = M_\nu^*\,.
\label{eq:ms}
\end{align}
Additionally, the consistent combination of flavour and gCP symmetries mandates that the matrix $X_\nu$ must satisfy the condition:
\begin{align}
X_\nu\, \rho^*(g_\nu)\, X_\nu^{-1} = \rho(g_\nu)\,.
\end{align}

Eqs.~\eqref{eq:ms} constrain the form of the unitary rotations diagonalizing the neutrino and charged-lepton mass terms, and therefore shape the Pontecorvo-Maki-Nakagawa-Sakata (PMNS) neutrino mixing matrix~\footnote{We consider the standard parametrisation of the PMNS, see e.g.~\cite{Olive:2016xmw}.},
which reads~\cite{Penedo:2017vtf}:
\begin{align}
U_{\rm PMNS} =  
P_e\, U_{23}(\theta^e,\delta^e)\, \Omega_e^\dagger\, 
\Omega_\nu\, R_{23}(\theta^\nu)\, P_\nu\, Q_\nu\,.
\label{eq:UPMNS}
\end{align}
Here, $P_{e,\nu}$ are permutation matrices, $Q_\nu = \text{diag}(1,i^{k_1},i^{k_2})$ with $k_{1,2} = 0,1$, and
\begin{align}
U_{23}(\theta^e,\delta^e) = 
\begin{pmatrix}
1 & 0 & 0 \\
0 &\cos\theta^e & \sin\theta^e\, e^{-i\delta^e} \\
0 & - \sin\theta^e\, e^{i\delta^e} & \cos\theta^e
\end{pmatrix},\,
R_{23}(\theta^\nu) = 
\begin{pmatrix}
1 & 0 & 0 \\
0 &\cos\theta^\nu & \sin\theta^\nu \\
0 & - \sin\theta^\nu & \cos\theta^\nu
\end{pmatrix}.
\end{align}
The matrices $\Omega_{e,\nu}$ are fixed by the choice of $G_f$ and of the specific residual symmetries. Thus, apart from discrete ambiguities, the PMNS matrix is determined by three real parameters: 2 angles, $\theta^{e}, \theta^{\nu} \in [0,\pi)$, and 1 phase, $\delta^e \in [0,2\pi)$.

\section{Application to $\boldsymbol{G_f = S_4}$}
$S_4$ is the non-Abelian symmetric group of permutations of four objects (e.g.~the vertices of a tetrahedron). It has 24 elements, admits 5 irreducible representations~\footnote{Our conclusions are independent of the choice of 3D representation.}, $\mathbf{1},\mathbf{1'},\mathbf{2},\mathbf{3},\mathbf{3'}$, and is conveniently described by the generators $S,T,U$, satisfying $S^2 = T^3 = U^2 = (ST)^3 = (SU)^2 = (TU)^2 = (STU)^4 = 1$. 

After identifying redundancies -- i.e.~some residual symmetries lead to the same PMNS -- and excluding phenomenological outliers (degenerate neutrino masses, texture zeros in the PMNS), we find 4 possible forms of the PMNS up to permutations of its rows and columns \cite{Penedo:2017vtf}:
\begin{align}
U_{\rm PMNS}^{\rm A} &= \begin{pmatrix}
\frac{1}{\sqrt{2}}\, e^{-{i\pi}/{6}} & \frac{1}{2}\,e^{i\,\star} & \frac{1}{2}\, e^{i\,\star} \\
\star & \star & \star \\
\star & \star & \star 
\end{pmatrix}\,,\,\,
U_{\rm PMNS}^{\rm B} = \begin{pmatrix}
\frac{1}{\sqrt{2}}\, e^{{i\pi}/{3}} & \star\,e^{i\pi/3} & \star\, e^{i\pi/3}\\
\star & \star & \star \nonumber \\
\star & \star & \star 
\end{pmatrix}\,,\,\,
\\
U_{\rm PMNS}^{\rm C} &= \begin{pmatrix}
\frac{1}{{2}}\, e^{{i\pi}/{3}} & \star\,e^{-i\pi/6} & \star\, e^{-i\pi/6}\\
\star\,e^{-i\pi/6} & \star & \star \\
\star & \star & \star 
\end{pmatrix}\,,\,\,
U_{\rm PMNS}^{\rm D} = \begin{pmatrix}
\frac{1}{{2}}\, e^{{i5\pi}/{6}} & \star & \star\\
\star\,e^{i\pi/3} & \star & \star \\
\star & \star & \star 
\end{pmatrix}\,,
\end{align}
where each star represents a different (and often lengthy~\cite{Penedo:2017vtf}) function of $\theta^e$, $\theta^\nu$, and $\delta^e$. The predicted absolute values and relative phases are instead written explicitly.

Not all permutations are viable if one takes into account the 3$\sigma$ ranges of 
neutrino oscillation parameters obtained in the global 
analysis of Ref.~\cite{Capozzi:2017ipn}. There are in fact 15 surviving viable cases,
which we denote A1, A2, B1-B4, C1-C5, D2-D5 \cite{Penedo:2017vtf}. Four of these (A1, A2, D4 and D5) are however strongly disfavoured by data ($\chi^2_\textrm{min} \gtrsim 15$).

\section{Results: Correlations and $\boldsymbol{0\nu\beta\beta}$-decay}
From the constraints on the form of the PMNS, correlations between angles
and phases follow.
For instance, in case C5, the following sum rule for $\cos \delta$ is satisfied:
\begin{align}
\cos\delta = \frac{1 - 4\cos^2\theta_{12} \sin^2\theta_{23} 
- 4\sin ^2\theta_{12} \cos^2\theta_{23} \sin^2 \theta_{13}}
{2\sin2\theta_{12} \sin2\theta_{23} \sin\theta_{13}}\,.
\end{align}
This case, for which $\chi^2_\textrm{min} = 0.5$, is obtained by permuting the rows and columns of $U_{\rm PMNS}^{\rm C}$ so that the fixed-magnitude element is brought to the 3-2 position.
Apart from cases B3, B4, and C1, all other cases satisfy a {\it bona fide} sum rule for $\cos \delta$~\footnote{i.e., with $\cos \delta$ given as a function of the $\theta_{ij}$, with no explicit dependence on $\theta^{e,\nu}, \delta^e$.}.

\begin{figure}
\centering
\includegraphics[height=8.5cm]{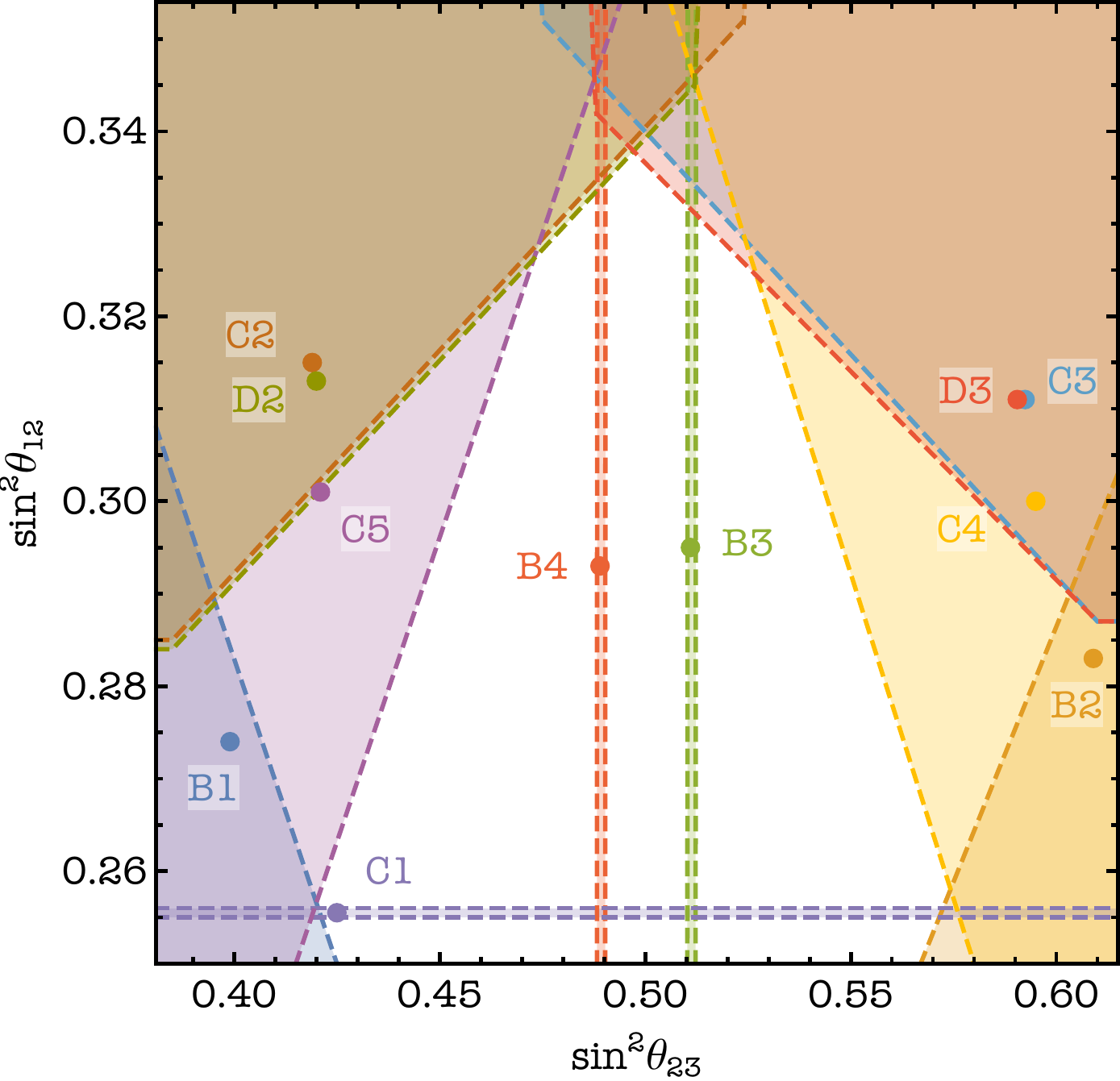}
\caption{Summary of symmetry constraints for viable cases with $\chi^2_\text{min} <15$, and for a neutrino spectrum with normal ordering (NO). Dots represent best-fit values for each case.}
\label{fig:all}
\end{figure}

For purposes of illustration, the constraints on the plane $(\sin^2 \theta_{23},\sin^2 \theta_{12})$ for each of the viable cases with $\chi^2_\text{min} <15$ are collected in Figure~\ref{fig:all}.

\begin{figure}
\centering
\includegraphics[height=7.2cm]{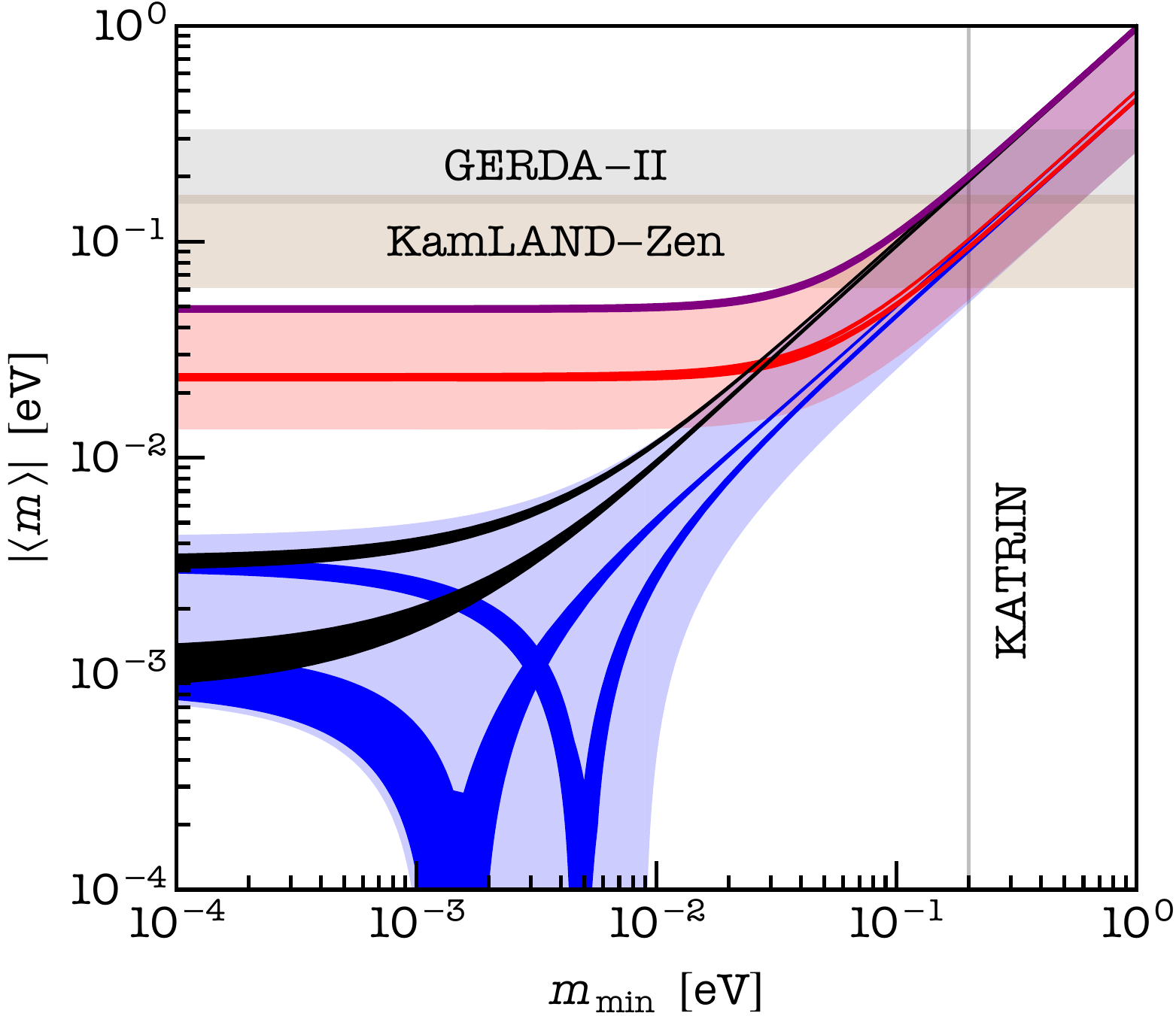}
\caption{Predictions for the effective Majorana mass in case C1 (see text) as a function of the smallest neutrino mass, $m_\text{min}$. Blue and red bands correspond to $k_1 = 0$, while black and purple bands refer to $k_1 = 1$. The KATRIN bound $m_\text{min} < 0.2$ eV is a prospective one.}
\label{fig:C1}
\end{figure}

Additionally, CPV phases are constrained to lie in specific intervals, and their values are strongly correlated. It is then possible to derive predictions for the neutrinoless double beta ($0\nu\beta\beta$)-decay effective Majorana
mass observable $|\langle m \rangle|$.
Case C1 in particular provides rather sharp predictions, since not only are mixing angles constrained but also CPV phases are fixed by symmetry to be $\alpha_{21} = k_1\, \pi$ and $\alpha_{31}- 2\delta = k_2\,\pi$. Predictions for $|\langle m \rangle|$ in this case are shown in Figure~\ref{fig:C1}.

\section{Summary and Outlook}
We have studied a class of models based on the breaking of $S_4 \rtimes H_\textrm{CP}$ flavour and generalised CP symmetry. We have found strong correlations between CPV phases and correlations between mixing angles and phases, including sum rules for $\cos \delta$ in some of the viable cases. We have additionally derived predictions for the $0\nu\beta\beta$-decay effective Majorana mass. Future data (Daya Bay, JUNO, T2K, T2HK, DUNE) will allow to test and discriminate between different symmetry predictions.
\pagebreak
\Acknowledgements
J.T.P. would like to thank the organisers of NuPhys2017 
for the opportunity to present this work.
This work was supported in part by the INFN
program on Theoretical Astroparticle Physics (TASP), 
by the PRIN 2012 project 2012CPPYP,
funded by the Italian Ministry of Education, University and Research (MIUR),
by the EU Horizon 2020 Marie Sk\l{}odowska-Curie grants 674896 and 690575, and by
the World Premier International Research Center Initiative (WPI
Initiative), MEXT, Japan (S.T.P.).


\begin{thebibliography}{99}
\setlength{\parskip}{0pt}

\bibitem{Olive:2016xmw}
  K.~Nakamura and S.~T.~Petcov in 
 C.~Patrignani {\it et al.} [Particle Data Group Collaboration],
  Chin.\ Phys.\ C {\bf 40} (2016) 100001.
    
\bibitem{Altarelli:2010gt}
  G.~Altarelli and F.~Feruglio,
  Rev.\ Mod.\ Phys.\  {\bf 82} (2010) 2701.

\bibitem{Ishimori:2010au}
  H.~Ishimori {\it et al.},
  Prog.\ Theor.\ Phys.\ Suppl.\  {\bf 183} (2010) 1.

\bibitem{Branco:1986gr}
  G.~C.~Branco, L.~Lavoura and M.~N.~Rebelo,
  Phys.\ Lett.\ B {\bf 180} (1986) 264.
 
\bibitem{Feruglio:2012cw}
  F.~Feruglio, C.~Hagedorn and R.~Ziegler,
  JHEP {\bf 1307} (2013) 027.

\bibitem{Holthausen:2012dk}
  M.~Holthausen, M.~Lindner and M.~A.~Schmidt,
  JHEP {\bf 1304} (2013) 122.
  
\bibitem{Penedo:2017vtf}
  J.~T.~Penedo, S.~T.~Petcov and A.~V.~Titov,
  JHEP {\bf 1712} (2017) 022.

\bibitem{Capozzi:2017ipn}
  F.~Capozzi {\it et al.}, 
  Phys.\ Rev.\ D {\bf 95} (2017) no.9,  096014.
\end{thebibliography}
\end{document}